\documentclass[journal=jacsat,manuscript=article]{achemso}

\usepackage[version=3]{mhchem} % Formula subscripts using \ce{}
\usepackage[T1]{fontenc}       % Use modern font encodings

\author{Jingwei Dong}
\affiliation[LSI]
{Laboratoire des Solides Irradiés, CEA/DRF/lRAMIS, CNRS, Ecole Polytechnique, Institut Polytechnique de Paris, 91128 Palaiseau, France}
\author{Dongbin Shin}
\affiliation [GIST]{Department of Physics and Photon Science, Gwangju Institute of Science and Technology (GIST), Gwangju 61005, Republic of Korea}
\affiliation[MP] 
{Max Planck Institute for the Structure and Dynamics of Matter and Center for Free-Electron Laser Science, Luruper Chaussee 149, 22761 Hamburg, Germany}
\author{Ernest Pastor}
\affiliation[INAM]
{University of Rennes, CNRS, IPR (Institut de Physique de Rennes) - UMR 6251, F-35000, Rennes, France}
\author{Tobias Ritschel}
\affiliation[TU DRESDEN]
{Institut f\"ur Festk\"orper- und Materialphysik, Technische Universit\"at Dresden, 01069, Dresden, Germany}
\author{Laurent Cario}
\affiliation[Versailles]
{Universit\'e de Nantes, CNRS, Institut des Mat\'eriaux Jean Rouxel, IMN, Nantes, F-44000, France}
\author{Zhesheng Chen}
\affiliation[LPS]
{Laboratoire de Physique des Solides, Université Paris-Saclay, CNRS, 91405 Orsay, France}
\author{Weiyan Qi}
\affiliation[LSI]
{Laboratoire des Solides Irradiés, CEA/DRF/lRAMIS, CNRS, Ecole Polytechnique, Institut Polytechnique de Paris, 91128 Palaiseau, France}
\author{Romain Grasset}
\affiliation[LSI]
{Laboratoire des Solides Irradiés, CEA/DRF/lRAMIS, CNRS, Ecole Polytechnique, Institut Polytechnique de Paris, 91128 Palaiseau, France}
\author{Marino Marsi}
\affiliation[LPS]
{Laboratoire de Physique des Solides, Université Paris-Saclay, CNRS, 91405 Orsay, France}
\author{Amina Taleb-Ibrahimi}
\affiliation[Soleil]
{Soci\'{e}t\'{e} civile Synchrotron SOLEIL, L'Orme des Merisiers, Saint-Aubin - BP 48, 91192 GIF-sur-YVETTE, France}
\author{Noejung Park}
\affiliation[NP] 
{Department of Physics, Ulsan National Institute of Science and Technology (UNIST), UNIST-gil 50, Ulsan 44919, Korea}
\author{Angel Rubio}
\affiliation[MP] 
{Max Planck Institute for the Structure and Dynamics of Matter and Center for Free-Electron Laser Science, Luruper Chaussee 149, 22761 Hamburg, Germany}
\author{Luca Perfetti}
\affiliation[LSI]
{Laboratoire des Solides Irradiés, CEA/DRF/lRAMIS, CNRS, Ecole Polytechnique, Institut Polytechnique de Paris, 91128 Palaiseau, France}
\email{luca.perfetti@polytechnique.edu}
\author{Evangelos Papalazarou}
\affiliation[LPS]
{Laboratoire de Physique des Solides, Université Paris-Saclay, CNRS, 91405 Orsay, France}

\title
 {Electronic dispersion, correlations and stacking in the photoexcited state of 1T-TaS$_2$}

\abbreviations{IR,NMR,UV}

\begin{document}

%% require t\textcolor[rgb]{0,0,0}{hat} it is printed as part of the abstract page. It will

\begin{abstract}

Here we perform angle and time-resolved photoelectron spectroscopy on the commensurate Charge Density Wave (CDW) phase of 1T-TaS$_2$. Data with different probe pulse polarization are employed to map the dispersion of electronic states below and above the chemical potential. Upon photoexcitation, the fluctuations of CDW order erase the band dispersion and squeeze the electronic states near to the chemical potential. This transient phase sets within half a period of the coherent lattice motion and is favored by strong electronic correlations. The experimental results are compared to Density-Functional Theory (DFT) calculations with a self-consistent evaluation of the Coulomb repulsion. Our simulations indicate that the screening of Coulomb repulsion depends on the stacking order of the TaS$_2$ layers. The entanglement of such degrees of freedom suggest that both the structural order and electronic repulsion are locally modified by the photoinduced CDW fluctuations.

\end{abstract}


The transition metal 1T-TaS$_2$ is a layered insulator with a rich phase diagram as a function of pressure and temperature. Its broken symmetry phases include the incommensurate, nearly commensurate, and Commensurate Charge Density Wave (C-CDW) \cite{Wilson}. Within each layer, the Ta lattice undergoes a periodic distortion in which 13 Ta ions form clusters with the motif of a Star-of-David (SD) \cite{Smaalen}. These clusters have an odd filling and lock-in to a C-CDW below 180 K. The observed insulating behavior of the C-CDW phase is generally attributed to the Mott localization of the electron in the highest occupied state of SDs \cite{Perfetti2005,Perfetti2003,Perfetti2005b}. A superconducting phase develops upon pressure \cite{Forro}, whereas metastable states can be reached by the application of laser \cite{Mihailovic2014} or current pulses \cite{Mihailovic2015,Cho}. These entwined orders emerge from the interplay of electron-phonon and electron-electron interactions, both being particularly strong in this dichalcogenide. 

Although widely believed to be a Mott insulator, the C-CDW phase also features an interlayer stacking with SDs dimerization. The stacking of two adjacent layers can be of three different kinds. Figure 1A show the top Aligned ($A$) and Laterally displaced ($L$) stacking with a vector of magnitude $2a$. The ground state dimerized geometry of TaS$_2$ is formed by alternating stacking between $A$ and $L$ configurations, called $AL$ stacking \cite{Lee}. The doubling of unit cell along the \textit{c}-axis direction has been confirmed by many different experiments, such as: X-Ray Diffraction (XRD) \cite{Wang}, Angle-Resolved Photoelectron Spectroscopy (ARPES) \cite{Ritschel2015,Ritschel2018,Lee}, Scanning Tunneling Microscopy (STM) \cite{Butler} and Low Energy Electron Diffraction (LEED) \cite{Hammer}.  %With this stacking configuration, two possible cleavage planes can terminate the surface, one leaving an intact bilayer at the surface and the other with an undimerized topmost layer \cite{Butler}.

By hosting an even number of electrons, the dimerized unit cell of the commensurate CDW cannot be a pure Mott phase. As in the case of VO$_2$ \cite{Biermann}, the instability of 1T-TaS$_2$ results from the interplay of strong correlations and structural distortion. This duality gave origin to several works, addressing the Slater-towards-Mott character of the ground state. Recent calculations revised the strength of Coulomb repulsion in this family of compounds and highlighted the strong effects that electronic interactions have on the band structure of 1T-TaS$_2$ \cite{Rubio}. Notable arguments backing this point of view are: the Mott gap observed in monolayer 1T-TaS$_2$ \cite{Monolayer}, the high sensitivity of the insulating state to non-isoelectronic substitution \cite{Fazekas}, and the dynamical response of electronic states upon photoexcitation \cite{Perfetti2006,Perfetti2008}. In particular, the ultrafast collapse of the gap at relatively low excitation density has often been availed as the major indication of strong electron-electron interaction \cite{Perfetti2006,Cavalleri,Rossnagel}.
Despite the large amount of experimental work on this subject, none of the time resolved data has been acquired with good energy and wavevector resolution.

This work reports time-resolved ARPES measurement on high-quality single crystals of 1T-TaS$_2$ in the insulating C-CDW phase. By making use of different polarizations of the probe pulse, it is possible to visualize the dispersion of electronic states below and above the chemical potential.  Moreover, time-resolved ARPES maps acquired with $S$ polarized probe disclose novel aspects of the photoinduced phase transition. The pump pulse erases the dispersion of the electronic states near to the chemical potential, and squeeze them into flat bands within half a period of the coherent CDW motion. Besides the oscillations of CDW amplitude, we propose that photoexcitation also engenders local variations of dimerization, orbital filling, and Coulomb repulsion. The combination of these effects triggers the melting of the Mott-Peierls gap. In order to gain more insights on this process, the electric states have been modeled by Density Functional Theory calculations with the Generalized Orbital $U$ (DFT+GOU) approach\cite{Rubio}. Our results show that the screened Coulomb repulsion depends on the stacking configuration. Such connection between structure and correlation should be carefully considered in any model aimed at simulating of the photoinduced CDW fluctuations.

\section{Data analysis and discussion}

Single crystals of 1T-TaS$_2$ have been grown by vapor transport in the form of shiny plaquettes. The batch of samples has been characterized by x-ray diffraction and transport measurements. Low-energy electron diffraction has been employed to verify the presence of the CDW reconstruction at the surface of the cleaved samples. Time-resolved ARPES experiments have been carried out on a single crystal cleaved at room temperature and subsequently cooled to 135 K, where all presented measurements have been performed. The sample is photoexcited by a pump pulse of 300 $\mu$J/cm$^2$ and centered at 1.55 eV \cite{Faure}. No sign of multiphoton emission of the pump beam could be detected at the measured fluence. Photoelectrons are emitted by a delayed pulse at 6.2 eV, with a bandwidth of 30 meV and a duration of 170 fs. This parameter choice is an optimal tradeoff between energy and temporal resolution for this specific experiment. The probe is focused on the sample on a spot of roughly $100\times 100$ microns. We reduced the photon flux of the 6.2 eV beam until no sign of space charge effects could be detected in the acquired photoelectron spectra. Note that photoelectrons emitted with 6.2 eV probe have an escape depth of roughly $3-5$ nm \cite{BP}, therefore detecting $\sim 5-8$ layers of 1T-TaS$_2$.

Figure 1B shows the unreconstructed Brillouin Zone (BZ) of primitive 1T-TaS$_2$ lattice and the reconstructed BZ of a periodic structure with C-CDW lattice modulation and $AL$ staking. In our setup (see Fig. 1C), the $P$ polarization corresponds to an electric field having equal projections along the $\overline{\Gamma}-\overline{M}$ and the $\textit{c}$-axis direction. In the following we will always refer to the polarization of the probe beam. Indeed the decoherence time of optical excitations, is too short to observe any dependence of the phoelectron intensity on the polarization of the pump. Figure 1D,E show photoelectron intensity maps acquired along the $\overline{\Gamma}-\overline{M}$ direction of the unreconstructed Surface Brillouin Zone (SBZ), with $P$ polarized probe and pump-probe delay of -200 fs (panel D) or 0 fs (panel E). An internal reference of pump-probe cross-correlation is obtained by monitoring the temporal evolution of states well above the chemical potential (see Fig. 1G).

Due to the interplay of electron-electron interaction and stacking order, we refer to the states below or above the chemical potential (zero of energy axis) as Lower Mott-Peierls Band (LMPB) and Upper Mott-Peierls band (UMPB), respectively. Since the UMPB also displays a sizable c-axis dispersion, the spectral weight redistribution of the ARPES intensity varies with probe photon energy \cite{Hofmann}. In the case of 6.2 eV probe photons, the LMPB peaks near -0.2 eV for $k_{||}=0.2$ \AA$^{-1}$~ and moves towards the chemical potential for $k_{||}\rightarrow 0$. This electronic structure is expected for $k_{\perp}$ projections where the energy distance between LMPB and UMPB becomes minimal \cite{Ritschel2018,Hofmann}. Figure 1F displays the temporal evolution of the photoelectron signal integrated around $\overline \Gamma$. Upon photoexcitation, the spectral weight is suddenly transferred to higher energy, inducing the ultrafast filling of the gap. A displacive excitation of SDs breathing mode provokes large and periodic modulations of the LMPB peak. The frequency of this symmetric mode corresponds to 2.4 THz, which is in agreement with the value extracted via stimulated Raman scattering \cite{Demsar}, electron diffraction \cite{Gedik2018} and previous time resolved ARPES results \cite{Perfetti2006,Cavalleri,Rossnagel}.

New insights into the structure of electronic states can be obtained by collecting photoelectron maps with $S$ polarized photons. As shown in the sketch of Fig. 1C, the electric field of $S$ polarized light lies in the surface plane and is perpendicular to the analyzer slits. The essential role played by the probe polarization on the ARPES maps of 1T-TaS$_2$ has been overlooked in previous experiments \cite{Perfetti2006,Perfetti2008,Cavalleri,Rossnagel,Bovensiepen} while it deserves special care in the data analysis. Depending on the lattice structure, a dichroic effect can indeed modulate the photoemission intensity of electronic states, either exalting or hindering the visibility of specific features. Similar observations have been recently done in black phosphorous, where the linear dichroism is discussed in a pseudospin representation \cite{Kim}. These effects are as drastic in 1T-TaS$_2$ as they are in black phosphorous (see the intensity maps at photon energy 96 eV and shown in the supplementary information file). As shown in Fig. 2A, the $S$ polarization strongly reduces the emission intensity of the LMPB while increasing the emission intensity of states above the chemical potential.  It is now possible to see that an UMPB is transiently occupied by the pump laser pulse. Our experimental observation is consistent with the orbital $d_{z^2}$ character of the LMPB and a mixed $d_{z^2}-d_{x^2-y^2}$ character of the UMPB (see supplementary information file). Indeed, the dipolar moment leading to photoelectron emission points along the $c$-axis for $d_{z^2}$ orbital while it lies in-plane for the $d_{x^2-y^2}$ one. Since the data of Fig. 2A have been acquired with $S$ polarized light, no component of the electric field can couple to the $d_{z^2}$ orbital, and the spectral weight of the UMPB follows the dispersion of bands with $d_{x^2-y^2}$ character. The $k$-space mapping of electronic states below and above the chemical potential allows for a precise estimate of the gap size: UMPB approaches LMPB towards the center of the SBZ, where their peak-to-peak energy distance becomes $0.15 \pm 0.1$ eV (blue line in Fig. 2A and Fig. 1D). This value is comparable to the gap obtained by means of infrared \cite{Infrared} and Raman spectroscopy  \cite{Raman}.

At a delay time of -200 fs the leading tail of the probe pulse has a small but finite overlap with the pump pulse (see the cross correlation of Fig. 1G). It follows that Fig. 2A shows a map that is still representative of a weakly photoexcited state. Subsequently, the absorbed energy density increases with pump-probe delay, leading to a full gap collapse. The snapshots in Fig. 2A-E indicate a complete restructuring of the electronic states, characterized by spectral weight transfer from the dispersive branches of the UMPB towards ingap states near the center of the SBZ. After 200 fs from the arrival of the pump pulse, the intensity map of Fig. 2E corresponds to states with low wavevector dispersion, due to the electronic localization in fluctuating nanodomains of the CDW \cite{Rossnagel}. This photoinduced flattening of electronic states dispersion during the CDW melting is one of the most relevant outcome of this article.

It is instructive to compare the dynamics of the melting process with the period of the CDW amplitude. Figure 3A shows the temporal evolution of the ARPES intensity integrated around $k_{||}=0.15$ \AA$^{-1}$. Periodic oscillations of the LMPB can be followed by plotting the ARPES intensity around -70 meV (see Fig. 3B). The coherent CDW motion displays the evolution expected from the displacive excitation of the SD breathing mode: it does take off near zero delays (more precisely -30 fs) and oscillates as a cosinus \cite{Papalazarou} with a period of 410 fs. As shown in Fig. 3A, the UMPB and LMPB approach to each other during the first half period of the CDW amplitude oscillation, nearly following the CDW displacement. This effect can be quantified in Fig. 3C, which shows energy distribution curves extracted from Fig. 3A for different pump-probe delays. At zero delay, the UMPB-LMPB distance is still close to the maximal value of 300 meV, while it reduces to 150 meV for a delay time of 150 fs.

Well before the reduction of gap magnitude, a midgap intensity grows up between LMPB and UMPB, reaching at zero delays already 80\% of the maximal value. The appearance of midgap states on a timescale faster than the coherent lattice motion has always been ascribed to Mott physics \cite{Perfetti2006,Perfetti2008,Cavalleri,Rossnagel,Bovensiepen}. Nonetheless, the sudden melting of the gap seems to be a general aspect of ultrafast phase transitions \cite{Gedik2019,Bauer,Brouet}. Here it is shown that the emergence of midgap states comes together with the flattening of electronic states dispersion, being an hallmark of CDW fluctuations. The disentanglement of electronic and structural degrees of freedom is impossible, as correlation effects, CDW order and stacking configuration are tightly connected.

This hypothesis is explored by calculations of the electronic states. The DFT is implemented in the Quantum ESPRESSO package using the PBE-type functional to approximate the exchange-correlation potential. Wave functions are obtained via the projector-augmented plane wave method and a basis set with a cutoff energy of 60 Ry. The lattice constant for bulk 1T-TaS$_2$ are $a=3.36$ \AA, $c=6.03$ \AA~ and a $3 \times 3 \times 6$ k-point mesh samples the Brillouin zone. In the case of DFT + GOU, the $\overline U$ potential of a SD cluster is obtained self consistently via the ACBN0 method, where $\overline U$ and $\overline J$ parameters are determined through the theory of screened Hartree-Fock exchange potential in a correlated subspace \cite{Rubio}. Remark that also other methods, such as the $GW-EDMFT$, have been recently able to account for correlation and dimerization of the 1T-TaS$_2$ groundstate \cite{Werner,Werner1}. Compared to $GW-EDMFT$ calculations, the $DFT+GOU$ method has both advantages and drawbacks: it does not retrieve the full spectral function but is fully \emph{ab-initio} (no adjustable parameters) and can treat the multiband problem without restricting the Hilbert space to one effective orbital. 

The band structure calculated by $DFT+GOU$ for $AL$ (Fig. 1A) and $L$ (Fig. 4A) stacking is shown in Fig. 4B. We limit the analysis to these two stacking configuration because they have been predicted to be the most stable and are very nearby in energy \cite{Lee}. Next, we show in Fig. 4C and 4D the unfolded spectral weight along the $\Gamma-M$ direction for $AL$ stacking and $L$ stacking, respectively. Our simulations indicate that electronic correlations strongly affect the electronic states, stabilizing the insulating phase. Figure 4E shows the direct gap $\Delta$ as a function of Coulomb repulsion $U$ for the case of dimerized $AL$ stacking and undimerized $L$ stacking. In the case of $AL$ stacking, the gap is $\Delta=0.1$ eV if correlations are ignored ($U=0$) while it increases up to $\Delta_{AL}=0.28$ eV when $U$ is equal to the self-consistent $\overline U_{AL}=0.45$ eV. Instead, for pure $L$ stacking the system is metallic for $U\leq 0.2$ eV and attains a gap $\Delta_{L}=0.12$ eV when $U$ is equal to the self consistent $\overline U_{L}=0.33$ eV. The self consistent $\overline U$ depends on the stacking order because of the different bandwidth along the $c$-axis, which weakens the effective screening when dimers are formed. Although our calculations are done in equilibrium conditions, the entanglement between CDW order, stacking and Coulomb repulsions must be important also in the photoexcited state.
Figure 4F plots the Density Of Electronic States (DOS) for $AL$ and $L$ stacking. The distance between the nearest peaks below and above the chemical potential (arrows on the figure) is roughly 30\% percent larger than the gap value, corresponding to 0.36 eV for the $AL$ case and 0.17 eV for the $L$ case. These values agree with the scanning tunneling spectroscopy spectra measured on the $A$, and $L$ termination of the 1T-TaS$_2$ surface \cite{Butler}. 

%This polarization and orbital dependence is not included in the spectral weight calculations in Fig. 1D,E) One step ARPES simulations would be instemad be necessary to account fo the dipole matrix element and polarization effects.

\section{Conclusions}

In conclusion, self-consistent DFT+GOU calculations predict that the C-CDW of 1T-TaS$_2$ is an insulator in which an out-of-plane dimerization and electronic correlations are highly entangled. New time resolved ARPES data acquired with $S$ polarization of the probe photons show that electronic states near to the chemical potential lose the wavevector dispersion during the CDW melting. This process indicates that photoinduced fluctuations partially disrupt the long-range order within half a period of the CDW amplitude mode. Time resolved model tracking on equal footing the disordered lattice motion, coherent CDW oscillations and screened Coulomb repulsion will be necessary to accurately reproduce the evolution of electronic states during the ultrafast phase transition.

\begin{acknowledgement}

The authors declare no Competing Financial or Non-financial interests. If required, the corresponding author and coauthors will make materials, data, code, and associated protocols promptly available to readers. 

Evangelous Papalazarou, Jingwei Dong, Ernest Pastor, Zhesheng Chen and Weiyan Qi performed the time resolved ARPES measurements and the data analysis. Dongbin Shin, Angel Rubio and Noejung Park have done the theoretical calculations. Tobias Ritschel, Marino Marsi and Amina Taleb-Ibrahimi participated to the scientific discussion. Laurent Cario has grown the 1T-TaS$_2$ single crystals and Romain Grasset characterized them via resistivity measurements. Luca Perfetti analyzed the data and wrote the article.

\end{acknowledgement}

%\begin{suppinfo}

%\begin{itemize}
%  \item Support information file: contains details on the sample characterization and on the experimental protocols.
%\end{itemize}

%\end{suppinfo}


\newpage

\begin{figure}
\includegraphics[width=\columnwidth]{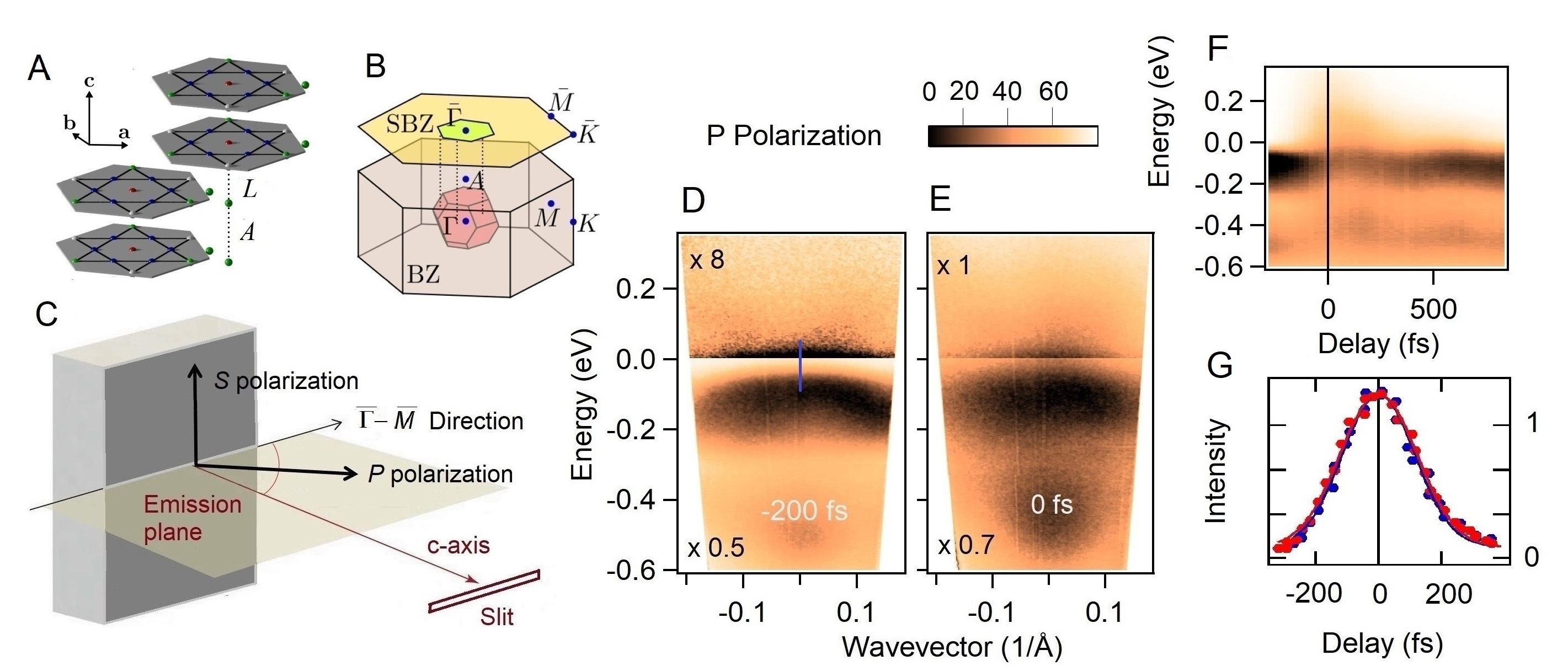}
\caption{ A) Unit cell of the $AL$ stacking in the C-CDW phase of 1T-TaS$_2$. B) Bulk Brillouin Zone (BZ) and surface Brillouin Zone (SBZ) of the undistorted and distorted structure of 1T-TaS$_2$. The high symmetry points $\Gamma$, $A$, $K$, $M$, $L$ refer to BZ of the undistorted structure and $\overline{\Gamma}$, $\overline{M}$, $\overline{K}$ are the respective projections on the SBZ. C) Orientation of the $P$ and $S$ polarization in the experimental geometry of the present work. The electric field of $P$ polarized light forms an angle of $45^\circ$ both with the $\overline{\Gamma}-\overline{M}$ direction and the $c$-axis direction. D-E) Photoelectron intensity map acquired with \textbf{\textit{P} probe polarization} along the $\overline{\Gamma}-\overline{M}$ direction for pump-probe delay of -200 fs (panel D) and 0 fs (panel E). The intensities above and below the Fermi level have been multiplied by rescaling factors in order to better visualize the electronic states with respect to a fixed color scale. The blue line in panel D stands for the electronic gap size. F) Photoelectron intensity acquired with $P$ probe polarization, integrated in the wavevector interval [-0.1,0.1] \AA$^{-1}$ and plotted as a function of pump-probe delay. G) Cross-correlation between pump and probe pulse obtained by extracting the photoelectron signal located 0.4 eV above the Fermi level. Blue and red circles correspond to $P$ and $S$ probe polarization, respectively.}
\label{Ppol}
\end{figure}

\begin{figure}
\includegraphics[width=\columnwidth]{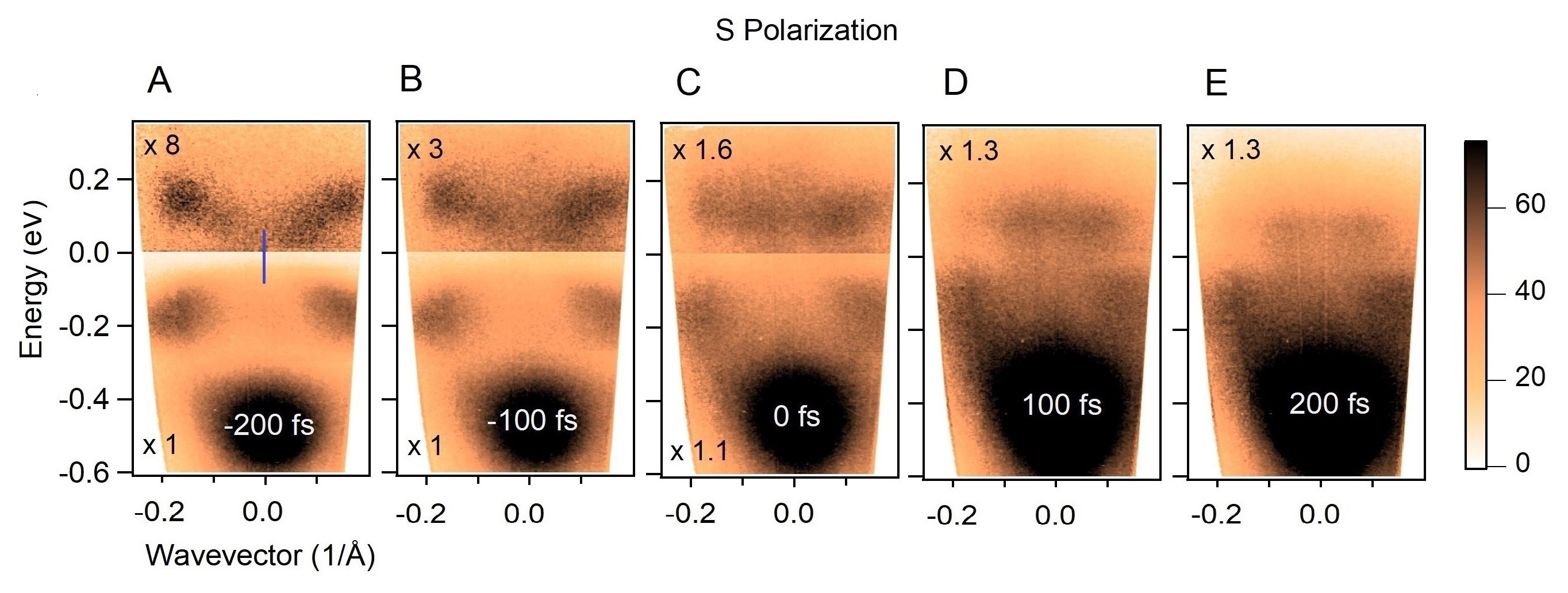}
\caption{ A-E) Photoelectron intensity map acquired with \textbf{\textit{S} probe polarization} along the $\overline{\Gamma}-\overline{M}$ direction for different pump-probe delays. The intensities above and below the Fermi level have been multiplied by rescaling factors in order to better visualize the electronic states with respect to a fixed color scale. The blue line in panel A stands for the electronic gap size.}
\label{Spol}
\end{figure}

\begin{figure}
\includegraphics[width=0.5\columnwidth]{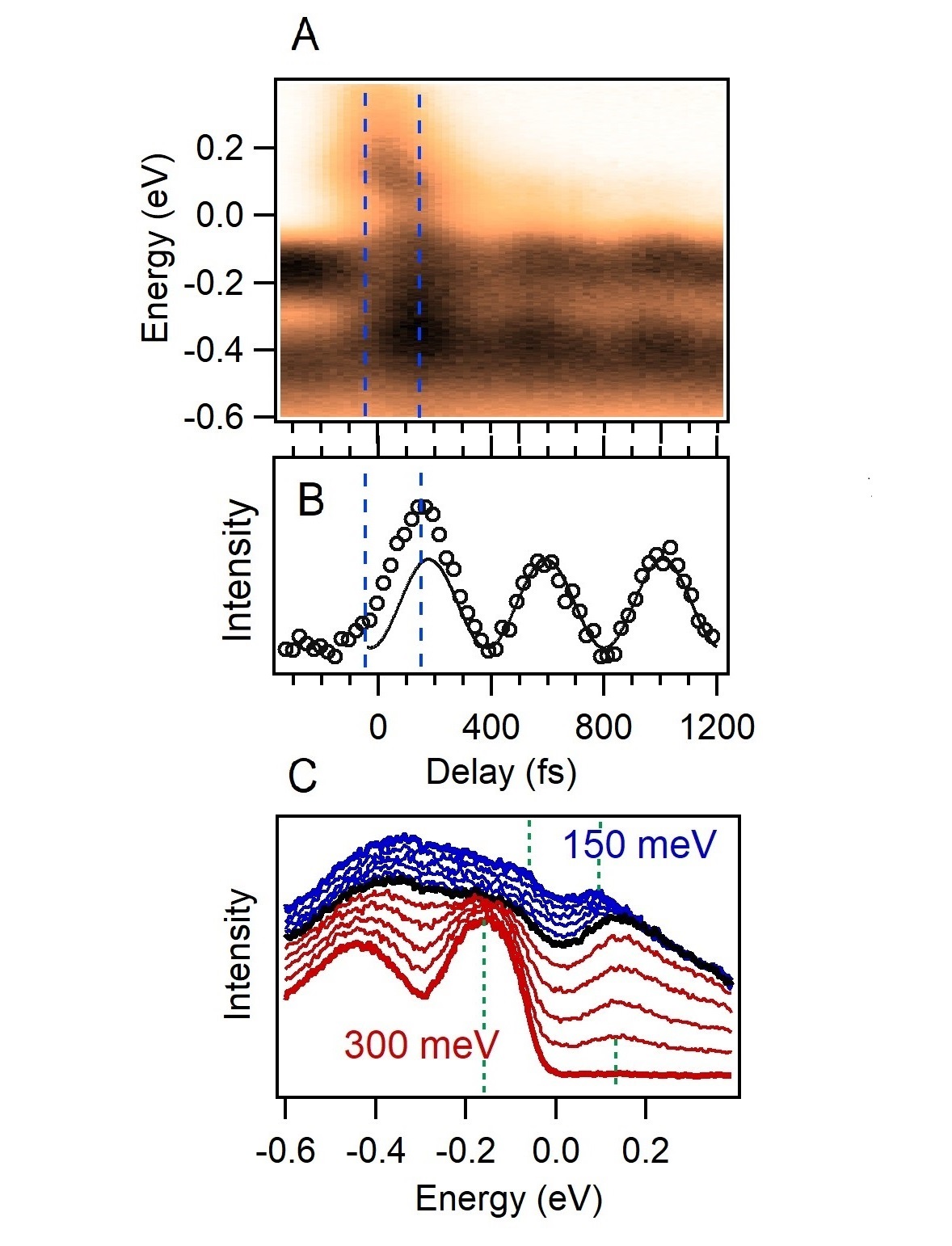}
\caption{A) Photoelectron intensity map acquired with \textbf{\textit{S} probe polarization} at wavevector $k=0.15$ \AA$^{-1}$~of the $\overline{\Gamma}-\overline{M}$ direction, as a function of pump-probe delay. B) Photoelectron intensity of panel A integrated with the energy interval [-65,-85] meV, which is dominated by coherent displacement of the CDW amplitude. The solid line is a cosinus fit of the oscillations for $\tau>400$ fs. C) Energy distribution curves extracted from the intensity map of panel A for pump-probe delays $-300<\tau<150$ fs. Red curves correspond to $-300<\tau<0$ fs, the black curve is at $\tau=0$, and blue curves are for $0<\tau<150$ fs.
}
\label{Dynamics}
\end{figure}

\begin{figure}
\includegraphics[width=\columnwidth]{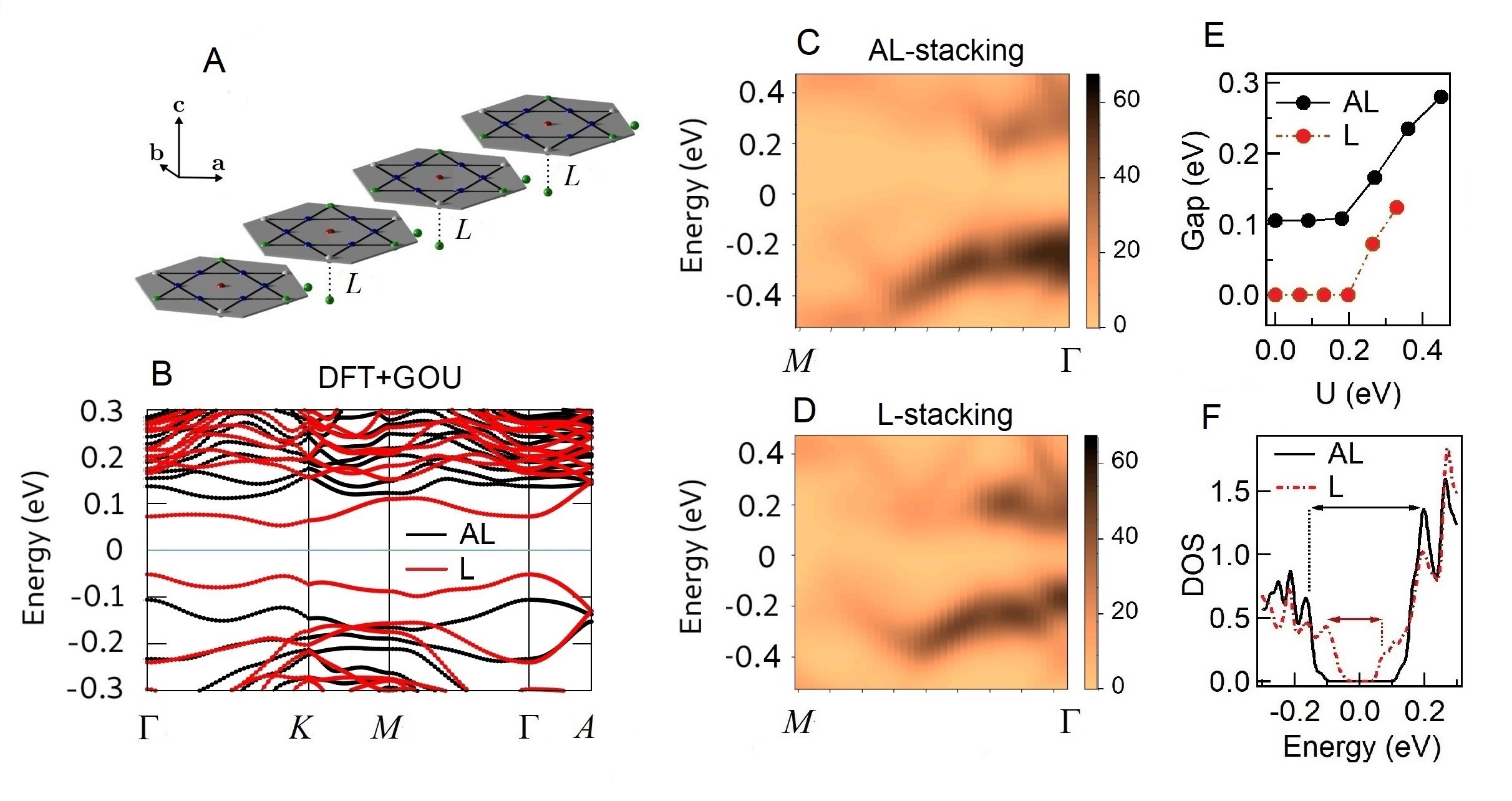}
\caption{A) Unit cell of the $L$ stacking in the commensurate CDW phase of 1T-TaS$_2$ B) Band structure of the $AL$ stacking (black lines) and $L$ stacking (red lines) calculated by DFT+GOU theory. The self-consistent $\overline U_{AL}=0.45$ eV and $\overline U_{L}=0.33$ eV are obtained via the ABCN0 method. C-D) Unfolded spectral weight distribution along the $\Gamma - M$ direction, calculated for the $AL$ and $L$ stacking. E) Electronic gap calculated by the DFT+GOU theory as a function of $U$ potential for the $AL$ and $L$ stacking. The largest $U$ value corresponds to the self-consistent finding $U=\overline U$. F) Density of electronic states of the $AL$ and $L$ stacking calculated by self-consistent DFT+GOU.}
\label{Theory}
\end{figure}

\end{document}